\documentclass{JHEP3} \usepackage{epsfig}

 \title{Measuring the Spin of Invisible Massive Graviton Excitations at
Future Linear Colliders}

 \author{Eri Asakawa\\ International Center for Elementary Particle Physics
         (ICEPP), University of Tokyo, and Theory Group, KEK\\
         E-mail: \email{eri@post.kek.jp}}
 \author{Kosuke Odagiri\footnote{Present affiliation:
         Institute of Physics, Academia Sinica, Nankang, Taipei,
         Taiwan 11529, The Republic of China}\\Theory Group, KEK\\
         E-mail: \email{odagirik@post.kek.jp}}
 \author{Yosuke Uehara\\Department of Physics, University of Tokyo\\
         E-mail: \email{yosuke@www-hep.phys.s.u-tokyo.ac.jp}}

 \abstract{
   We consider the production process $e^-e^+\to G\gamma$ of invisible
gravitons ($G$) at future linear colliders. We discuss whether the angular
distribution of the photon ($\gamma$) can be used to measure the spin of
the invisible graviton, or of any other invisible objects produced.
   We propose a method based on the Fourier expansion of the transverse
energy squared moment distribution of the photon. We provide justification
for this method, and confirm, especially for the case of two extra
dimensions, that the method is valid within a realistic setup, which
includes the simulation of the Standard Model background, beamstrahlung,
bremsstrahlung, calorimeter resolution and calorimeter coverage.
   When the number of extra dimensions is increased, the angular
distribution does not provide sufficient information to extract the spin,
but this method still offers a useful parameterization of the single photon
cross section using which the nature of the missing object can be studied.}


 \preprint{UT--ICEPP--02--04\\KEK--TH--820\\UT--02--20}

\begin{document}

%
%
%
%

 \section{Introduction}\label{introduction}

 \subsection{Theory Background}

The Standard Model (SM) of Particle Physics, which has so far stood the
test of numerous precision measurements, is nevertheless not considered to
be the ultimate theory. One of the reasons is a serious shortcoming of
the SM, namely that the mass of the Higgs boson diverges quadratically as
we evolve the energy scale of the theory towards the electroweak scale from
the Planck scale. This is the so-called Hierarchy Problem.

We expect that the SM as an effective theory is valid up to
$\mathcal{O}$(1\,TeV), and any signs of New Physics, if they should exist,
will arise at that energy scale.
  Supersymmetry is a well-motivated theory which solves the Hierarchy
Problem, but another possibility has been proposed \cite{EXTRADIMENSION}
recently. According to this theory, the SM particles are confined in a
3-brane, and only gravity propagates in the compactified extra dimensions.
The weakness of gravity is explained by the fact that it propagates in the
extra dimensions, and the overlap of the wavefunctions between the graviton
and the SM particles is small. The fundamental scale of the theory is not
$M_{pl} \sim 2.4 \times 10^{18}$ GeV, but $M_{D}=\mathcal{O}$(1\,TeV).
Here:
 \begin{eqnarray} M_{pl}^2 \sim V_{n} M_{D}^{n+2}, \end{eqnarray}
 where $n$ is the number of extra dimensions and $V_{n}$ is the volume of
the extra dimensions. Thus the weak scale and the fundamental scale $M_{D}$
are very close, and so the Hierarchy Problem is `resolved'. It may be
argued that this `resolution' only translates one problem, namely the
hierarchy between the electroweak scale and the fundamental scale, into
another, namely the size of the extra dimensions. Let us nevertheless adopt
this model as an effective theory applicable at low energies, leading to
concrete predictions which should be studied.

Since the graviton propagates in the compactified extra dimensions, there
are many massive Kaluza-Klein excitations. Such excitations couple to the
SM particles, and we can expect to observe at future colliders massive
graviton excitations, which we call massive gravitons hereafter for
simplicity.

Some bounds are available already from current collider experiments. D0
\cite{D0} and L3 \cite{L3} have made searches both in the direct
production channel where the produced graviton is undetected and is
identified through the missing 4-momenta, and in the indirect channel
where virtual graviton contributions give rise to `contact term' type
effective interactions. The direct limit on $M_{D}$ can be set only in the
former of these two search modes, and the current limit from L3 \cite{L3}
is $M_{D} \gtrsim$ 1 TeV for two extra dimensions.

When the number of extra dimensions is low, especially when $n=2$,
Astrophysics sets a stringent bound on the fundamental scale $M_{D}$.
Graviton emission into extra dimensions from a hot supernova core is
considered in Ref.~\cite{SUPERNOVA}, and the phenomenology of SN1987A
places a strong constraint on this energy loss mechanism, allowing us to
derive a bound on the fundamental scale $M_{D}$. When $n=2$, the allowed
lower bound on the fundamental scale $M_{D}$ is $M_{D} \gtrsim$ 50 TeV.
Furthermore, the graviton decay contribution to the cosmic diffuse gamma
ray radiation sets the bound $M_{D} \gtrsim 110$ TeV for the case $n=2$
\cite{HALL-SMITH}.
  The bound becomes much less stringent if we raise the number of extra
dimensions $n$. The constraint from SN1987A becomes $M_{D} \gtrsim 4$ TeV
for $n=3$, and $M_{D} \gtrsim 1$ TeV for $n=4$. There is theoretical
uncertainty associated with these numbers, and so the TeV fundamental
scale is not ruled out.
  In any case it is meaningful to derive alternative limits and so we do
not limit ourselves to parameter regions not excluded by Astrophysics.

  General studies of graviton measurement have been made in
\cite{COLLIDER1,Han:1998sg,COLLIDER2}. Search strategies for future $e^{+}
e^{-}$ colliders \cite{EECOLLIDER,NEARWORK}, the LHC \cite{PPCOLLIDER}, the
$\gamma\gamma$ colliders \cite{AACOLLIDER}, and $e\gamma$ colliders
\cite{EACOLLIDER} have also been studied.
  Recently, there has been increased attention on the higher energy
blackhole and `trans-Planckian' physics which may be observed at LHC
\cite{BLACKHOLE}.
  We do not consider these effects as the energy scale is lower in our
analysis, and hence the leading extra dimensional effects, in the form of
Kaluza-Klein excitations of the graviton, are expected to be
perturbatively calculable.

 \subsection{Motivation}

  We note that the above studies have concentrated mainly on the discovery
of the extra dimensional phenomenology. In this regard, although the
difference between the signal and the SM background has certainly been
considered and exploited in the above studies
\cite{COLLIDER1}--\cite{EACOLLIDER}, no systematic study has been
available so far, for the case of these `invisible' gravitons, to claim
the discovery of spin-2 massive excitations.

  Let us discuss this point further. For the case of massive spin-2
\emph{resonances}, the spin-2 nature is probed by measuring the decay angle
$\theta$. The presence of terms proportional to $\cos^4\!\theta$ then
provides evidence for the spin-2 nature of the excitation.

This same measurement can be made with the angular distribution resulting
from virtual graviton exchange. One may measure processes such as
$e^-e^+\to \gamma\gamma$ and the presence of terms proportional to
$\cos^4\!\theta$ can be confirmed in principle. This, however, is not
conclusive evidence of spin-2 gravitons, as all that one is measuring here
is the nature of the `contact-term' type interactions, ${\mathcal L}\propto
{T_\mathrm{SM}}^{\mu\nu}{T_\mathrm{SM}}_{\mu\nu}$, which arise from
integrating out virtual graviton contributions.

In order to claim the discovery of spin-2 massive excitations it is 
necessary, therefore, to produce them directly, and measure distributions 
associated with them. This is the topic of our present exposition.

  As each of the massive graviton modes has a coupling suppressed by
$M_{pl}$, it escapes detection and the events are characterized by missing
4-momenta. At hadron colliders, only the missing transverse momentum is
measurable, and so the extraction of the spin is, even if this is possible
at all, expected to be highly model dependent.
  Let us therefore concentrate on the case of lepton collisions, more
specifically the $e^-e^+$ future linear colliders.

At first sight, it may appear that the measurement of polarizations
participating in the interaction is sufficient. This is unfortunately not
so. For $e^-e^+$ annihilation the polarization structure is exactly the
same for both spin-1 and spin-2 objects. For $\gamma\gamma$ annihilation
the anticorrelation of the photon helicities is an interesting property of
the direct production of spin-2 objects, but it is not possible to trigger
such an event as no measurable particle is produced. Even in $e\gamma$
collisions it is not possible to utilize only the polarizations to
determine the spin of the excitation.

This leads us to the consideration of the process $e^-e^+ \rightarrow
G\gamma$ as the only way to measure the `spin' of the graviton excitation.

There are two variables which we can measure. These are the energy and the
polar angle of the photon. The energy of the photon measures the mass of
the excitation. A study is available in \cite{NEARWORK} which compares the
distribution of graviton events against that of the supersymmetric
$e^-e^+\rightarrow \widetilde{\chi}^0_1\widetilde{\chi}^0_1\gamma$. Our
approach is more systematic and rigourous.

Our starting point is the realization that the collinear divergence, which
dominates the $G \gamma$ cross section, is only logarithmic. At the
amplitude squared level, the divergence scales as $1/p_T^2$ where $p_T$,
or $E_T$ which is the quantity that is actually measured, is the photon
transverse momentum. Thus it is possible to regularize the collinear
divergence of any $\gamma+$missing events by multiplying the distribution
by $p_T^2$ or $E_T^2$.

The resulting distribution, which we may call the $E_T^2$ moment
distribution, shows a remarkable behaviour. In short, spin-$S$ excitations
have moment distributions which contain terms up to $\cos^{2S}\!\theta$,
or equivalently $\cos(2S\theta)$. Hence by Fourier analysis, in principle,
we can extract the nature of the excitation.

The SM background process which is at the leading order in the electroweak
coupling constant, $e^{-} e^{+} \rightarrow Z^0 \gamma \rightarrow \nu
\bar{\nu} \gamma$, only contains terms up to $\cos2\theta$. Furthermore,
we can reduce this background by cuts around the corresponding photon
energy.

The case of the subleading order background process, $e^-e^+\rightarrow
\nu_e\bar\nu_e\gamma$, which occurs both via virtual $W^\pm$ exchange and
via the off-shell $Z^0$, is less straightforward. We also calculate this
background. The $W^\pm$ induced contribution can contain terms
proportional to $\cos4\theta$ and higher. Hence this background must be
subtracted from the measured distribution.

As the analysis relies on measuring only one visible object, it is
necessary to carefully consider the systematics which can affect the
measurement. In addition to the calorimeter resolution and the angular
coverage, there are further obstacles from initial state radiation (ISR),
composed of bremsstrahlung and beamstrahlung.

We concentrate on analysing the systematics in this paper, and leave the
statistical analysis and the error estimation, as well as a more detailed
account of the hard/multiple photon emission effect, as topics for future
studies. In this regard, we should note that the analysis presented in
this paper is realistic, but the numbers are not conclusive.

We finally note that the work presented in this paper provides a general
framework for analysing events composed of $\gamma+$missing 4-momenta,
which can be used to search for any New Physics, not only those associated
with extra spacial dimensions. As an example we discuss the case of the
supersymmetric $\widetilde{\chi}^0_1\widetilde{\chi}^0_1\gamma$ process.

This paper is organized as follows: in Sect.~\ref{method} we discuss our
approach and explain the procedures adopted to make realistic simulation.
In Sect.~\ref{results} we present the results of our calculation. The
conclusions are stated at the end in Sect.~\ref{conclusions}.

 \section{Method and Discussions}\label{method}

 \subsection{Fourier Analysis}

  Let us first define the kinematic variables $\theta$ and $x_\gamma$.
$\theta$ is the angle between the photon and the momentum direction of the 
incoming electron. There is no forward--backward asymmetry and so our 
results are unchanged if we measure the angle from the positron direction.
The energy fraction is defined by:
 \begin{equation}
 x_\gamma = \frac{2E_\gamma}{\sqrt{s_\mathrm{nominal}}}. \label{xgamma}
 \end{equation}
  In the following, $s\equiv s_\mathrm{nominal}$ denotes the nominal
centre-of-mass energy squared. The effective centre-of-mass energy squared
is written as $\hat s$.
  The doubly differential distribution, $d^2\sigma/d\cos\theta dx_\gamma$,
has both soft and collinear divergences, and therefore we expect that
there is little visible distinction between the signal distribution and
the background distribution.
  For the sake of clarification, for the two-body case, disregarding the
effect of ISR, we have:
 \begin{equation}
 \frac{d^2\sigma}{d\cos\theta dx_\gamma} =
 \frac{x_\gamma}{32\pi s}\overline{|\mathcal{M}|^2},
 \end{equation}
  where $\overline{|\mathcal{M}|^2}$ is the corresponding matrix element 
squared summed over final state helicities and averaged over initial state 
helicities, and for the three-body case we have:
 \begin{equation}
 \frac{d^2\sigma}{d\cos\theta dx_\gamma} =
 \frac{x_\gamma}{512\pi^3}\int\frac{d\Omega^*}{4\pi}
 \overline{|\mathcal{M}|^2}.
 \end{equation}
  The integration in the three-body case is the integration over the solid
angle of the invisible `decay' in the rest frame of the `decaying' system.

  As stated in the introduction, this divergence is only logarithmic, and
so we can eliminate this divergence by multiplying the distribution by the
transverse energy squared, $E_T^2$, of the photon, viz:
 \begin{equation}
 \frac{d^2<\sigma E_T^2>}{d\cos\theta dx_\gamma} =
 \frac{E_T^2d^2\sigma}{d\cos\theta dx_\gamma}.
 \end{equation}
  We note that an alternative definition, which differs from the above
definition by factor $E_\gamma$, may be more useful in experimental
analyses:
 \begin{equation}
 \frac{d^2<\sigma E_T>}{d\theta dx_\gamma} =
 \frac{E_Td^2\sigma}{d\theta dx_\gamma}.
 \end{equation}
  We take the first definition as we consider it theoretically more
elegant.

  This $E_T^2$ moment distribution has the empirical property, which we
have verified analytically for several cases and we attempt to justify in
the next Subsection, that only angular dependencies up to
$\cos^{2S}\!\theta$ are present for the production of invisible spin-$S$
objects. Assuming this property, and noting that $\cos^\alpha\!\theta$ can
be expressed in terms of a sum of terms proportional to $\cos\beta\theta$,
$\beta\leq\alpha$, we can apply the Fourier expansion to this moment
distribution.
 \begin{eqnarray}
 \frac{E_T^2d^2\sigma}{d\cos\theta dx_\gamma} &=&
 a_0(x_\gamma)+a_1(x_\gamma)\cos2\theta+a_2(x_\gamma)\cos4\theta\ldots
 \nonumber\\&=& \sum_{m=0}^Sa_m(x_\gamma)\cos2m\theta.
 \end{eqnarray}
  The coefficients $a_m$ are obtained, at least in principle, as:
 \begin{eqnarray}
 a_0(x_\gamma) &=& \frac1{2\pi}\int_0^{2\pi}
 \frac{E_T^2d^2\sigma}{d\cos\theta dx_\gamma}
 d(2\theta), \label{a0} \\
 a_m(x_\gamma) &=& \frac1\pi\int_0^{2\pi}
 \frac{E_T^2d^2\sigma}{d\cos\theta dx_\gamma}
 \cos2m\theta d(2\theta) \qquad(m=1,2,3,\ldots).
 \label{an}
 \end{eqnarray}
  Hence the measurement of the maximum $m$ for which $a_m$ is nonvanishing
gives a measure of the `spin' of the produced object. We note that the
$a_m$'s are dimensionless quantities, which we measure in units of
fb\,GeV$^{2}$. Fourier analysis, as presented here, is not the only way to
extract these coefficients, but it is the most theoretically elegant. It
can also be considered, at the very least, as a way of parameterizing the
two dimensional distribution of $x_\gamma$ and $\cos\theta$.

  In the presence of ISR, when evaluating the Fourier coefficients, we need
to be more careful as to which frame $x_\gamma$ and $\cos\theta$ are
measured in. To be consistent, these should be evaluated in the lab frame
throughout. However, it is more convenient to carry out the calculation in
the centre-of-mass frame using the Monte-Carlo approach. In this case, one
must be careful to include the correct Jacobian factors.

  We have tabulated the leading order theoretical values for these
coefficients in Tabs.~\ref{tab_prefac} and \ref{tab_coeffs}. The
coefficients are obtained by multiplying the functions in
Tab.~\ref{tab_coeffs} by the prefactors in Tab.~\ref{tab_prefac} and
$\hbar^2=0.389\times10^{12}$ fb\,GeV$^2$. Further discussions are given in
the Subsections to follow.

 \begin{table}[ht]\begin{center}\begin{tabular}{|c|c|}\hline
 \rule[-0.35cm]{0cm}{0.9cm} &
 prefactor \\\hline
 graviton \rule[-0.35cm]{0cm}{0.9cm} &
 $\frac{\alpha_\mathrm{EM}}{1024}S_{n-1}\left(
 \frac{\sqrt{s}}{M_D}\right)^{n+2}
 x_\gamma(1-x_\gamma)^{n/2-1}$ \\\hline
 $Z^0\to\nu\bar\nu$ \rule[-0.35cm]{0cm}{0.9cm} &
 $\frac{\alpha_\mathrm{EM}^3(1-2\cos2\theta_W+2\cos^2\!2\theta_W)}
  {16\sin^4\!2\theta_W}\frac{x_\gamma(1-x_\gamma)}
  {\left(1-M_Z^2/s-x_\gamma\right)^2+M_Z^2\Gamma_Z^2/s^2}$
 \\\hline
 $\nu_e\bar\nu_e$ \rule[-0.35cm]{0cm}{0.9cm} &
 $\sim\frac{\alpha_\mathrm{EM}^3}{192\sin^4\!\theta_W}
 \frac{x_\gamma(1-x_\gamma)s^2}{M_W^4}$ \\\hline
 spin 1 \rule[-0.35cm]{0cm}{0.9cm} &
 $\propto x_\gamma(1-x_\gamma)\frac{s^2}{M^4}$ \\\hline
 spin 0 \rule[-0.35cm]{0cm}{0.9cm} &
 $\propto x_\gamma(1-x_\gamma)\frac{s^2}{M^4}$ \\\hline
 $\widetilde{\gamma}\widetilde{\gamma}$ \rule[-0.35cm]{0cm}{0.9cm} &
 $\frac{\alpha_\mathrm{EM}^3}{48}x_\gamma(1-x_\gamma)
 \left(\frac{s^2}{M^4_{\tilde{e}_L}}
 +\frac{s^2}{M^4_{\tilde{e}_R}}\right)
 \left(1-\frac{2M^2_{\tilde{\gamma}}}{s(1-x_\gamma)}
 \right)\sqrt{1-\frac{4M^2_{\tilde{\gamma}}}{s(1-x_\gamma)}}$\\\hline
 \end{tabular}
 \caption{Overall prefactor functions for the Fourier expansion. One
should also multiply by $\hbar^2=0.389\times10^{12}$ fb\,GeV$^2$.  
$\nu_e\bar\nu_e$ in the above table refers to the $W^\pm$ exchange, in
the limit of zero momentum transfer along the virtual $W^\pm$ propagator.
`spin 1' and `spin 0' tabulated above are for the production of heavy
bosons followed by the decay to two massless particles. The photino pair
production corresponds to the limit of heavy selectrons.}\label{tab_prefac}
 \end{center}
 \end{table}
 \begin{table}
 \begin{center}
 \begin{tabular}{|c|c|c|c|}\hline
 \rule[-0.2cm]{0cm}{0.6cm} &
 $a_0$ &
 $a_1$ &
 $a_2$ \\\hline
 graviton \rule[-0.2cm]{0cm}{0.6cm} &
 $32-64x_\gamma+60x_\gamma^2-28x_\gamma^3+5x_\gamma^4$ &
 $-12x_\gamma^2+12x_\gamma^3-4x_\gamma^4$ &
 $-x_\gamma^4$ \\\hline
 $Z^0\to\nu\bar\nu$ \rule[-0.2cm]{0cm}{0.6cm} &
 $8-8x_\gamma+3x_\gamma^2$ &
 $x_\gamma^2$ &
 $0$ \\\hline
 $\nu_e\bar\nu_e$ \rule[-0.2cm]{0cm}{0.6cm} &
 $\sim 8-8x_\gamma+3x_\gamma^2$ &
 $\sim x_\gamma^2$ &
 $\sim 0$ \\\hline
 spin 1 \rule[-0.2cm]{0cm}{0.6cm} &
 $8-8x_\gamma+3x_\gamma^2$ &
 $x_\gamma^2$ &
 $0$ \\\hline
 spin 0 \rule[-0.2cm]{0cm}{0.6cm} &
 $1$ &
 $0$ &
 $0$ \\\hline
 $\widetilde{\gamma}\widetilde{\gamma}$ \rule[-0.2cm]{0cm}{0.6cm} &
 $8-8x_\gamma+3x_\gamma^2$ &
 $x_\gamma^2$ &
 $0$ \\\hline
 \end{tabular}\end{center}
 \caption{Coefficient functions for the Fourier expansion.}\label{tab_coeffs}
 \end{table}

 \subsection{Spin and Angular Distribution} \label{distribution}

  The naive technical reason for the correlation between the spin $S$ and
the number of coefficients $m_\mathrm{max}$ of the Fourier expansion is as
follows.

  A spin $S$ boson is a rank $S$ tensor with $S$ indices. Each of these
indices, when contracted with the electron current or a momentum operator,
gives a contribution to the amplitude which is proportional to up to
$\cos\theta$. Hence we get $\cos^{2S}\!\theta$ at the amplitude squared
level.

  Perhaps a more physical restatement is that the plane wave due to the
produced spin-$S$, or rank $S$ tensorial, object intersects the electron
current $S$ times at angle $\theta$.


  There is a loophole in this argument concerning the case of the
$s$-channel photon mediated contributions. In this case the intermediate
photon, which is a vector, carries the angular information and so the
distribution is proportional to up to $\cos^2\!\theta$ already before
multiplying by $E_T^2$. Hence the moment distribution $(E_T^2d^2\sigma/
d\cos\theta dx_\gamma)$ contains terms up to $\cos^4\!\theta$, regardless
of the spin of the produced object.

  This is a limitation of our method, but there are only few cases where
this becomes problematic. If the $s$-channel graph is solely responsible
for the production, the collinear singularity is absent after background
subtraction, and so it becomes immediately clear that there is no
$t$-channel contribution. The ambiguity concerning the `spin' of the
produced object arises only in cases where the $s$-channel and $t$-channel
terms co-exist, and only for spin lower than 2, i.e., spin-0 or spin-1.

  Out of these two cases, the spin-1 case is genuinely problematic and the
signal is not distinguishable from a spin-2 signal. However, we note that
a photon--photon--vector vertex lacks theoretical motivation.

  If the produced object is spin-0, the $s$-channel term is
distinguishable from the $t$-channel term by beam polarization. Hence this
case is still not serious.

  Having explained this, it is instructive to now look more closely at
spin-2 production, as this case involves both the $s$-channel, $t$-channel
and contact terms.

  Out of these, the $s$-channel term is separately gauge invariant with
respect to the photon, and so is the sum of the $t$-channel and contact
terms. We need to consider whether the $\cos^4\!\theta$ behaviour
originates in the latter of these contributions. Indeed, in the Feynman
gauge, raising the tensor rank of the field only amounts to further
contractions with the metric tensor $g_{\mu\nu}$, and so by dimension
counting, and by forward--backward symmetry, we see that the contribution
in the spin-2 case only goes up to $\cos^2\theta$.

  On the other hand, the longitudinal contributions from terms that are
proportional to $p_\mu p_\nu$ do have the right dimensions for a
$\cos^4\!\theta$ behaviour, but we generally expect that such
contributions, coming from Goldstone modes, behave like objects with lower
spin.

  Explicit evaluation shows that such contributions do lead to
$\cos^4\!\theta$ behaviour. However, this behaviour comes from an
unphysical source, namely the non-conser\-vation of the energy--momentum
tensor. The inclusion of the $s$-channel graph is necessary. This cancels
the unphysical $\cos^4\!\theta$ behaviour and leads to the final
distribution which is physical and contains terms up to $\cos^4\!\theta$.

  It is possible to restate the preceding discussion in terms of the
individual helicity components. However, in our opinion, this type of
analysis only provides another viewpoint and does not enrich our insight
into the nature of the problem at hand. Although it is clear that we have
not provided a rigourous justification, to some extent, we believe that
this is a price that we have to pay for retaining generality.

  From the above discussion, it is clear that the presence of a tensorial
interaction is not sufficient to guarantee a $\cos^4\!\theta$ behaviour.  
Indeed, we have verified this for the case of an anomalous coupling vector
boson $V$, whose coupling to the electron is proportional to
$(1/\Lambda)\sigma_{\mu\nu}F^{\mu\nu}$. We have:
 \begin{equation}
 \overline{|\mathcal{M}|^2}(e^-e^+\to V\gamma)
 = \left(\frac{e}{\Lambda}\right)^2
 \left[4\left(s-m_V^2\right)+\frac{m_V^2}{ut}\left(s^2+m_V^4\right)
 \right],
 \end{equation}
 and so this interaction only gives rise to a $\cos^2\!\theta$ behaviour of
the $E_T^2$ moment distribution.

  What of the spin-3 and higher spin objects? Without explicit example it
is difficult to say, but we expect the above argument to hold, such that
the $\cos^6\!\theta$ term can originate solely from an unphysical origin.

  The case where the produced object is a composite system is discussed 
next.

 \subsection{Background and SUSY Signal}\label{sect_bkgd}

\begin{figure}[ht]
 \centerline{\epsfig{file=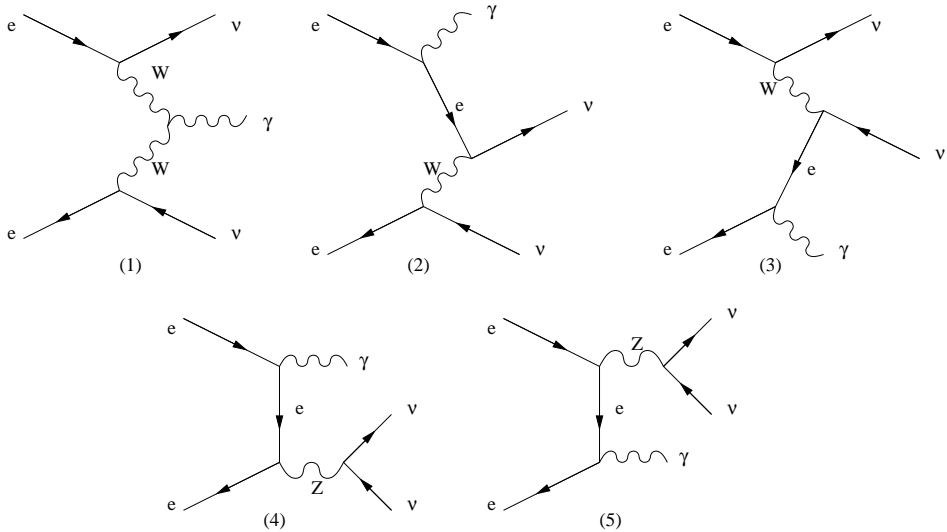,width=13cm}}
 \caption{The Feynman diagrams for the 
 $e^-e^+\to\nu\bar\nu\gamma$ process. Only the bottom two diagrams
contribute to the $\nu_\mu\bar\nu_\mu\gamma$ and
$\nu_\tau\bar\nu_\tau\gamma$ final states.}
 \label{nunuAfeyn}
 \end{figure}

  The leading Standard Model background comes from $\nu\bar\nu\gamma$,
which includes the contribution $Z^0\to\nu\bar\nu$, as well as the
$t$-channel $W^\pm$--exchange leading to the final state $\nu_e\bar\nu_e
\gamma$. The Feynman diagrams are as shown in Fig.~\ref{nunuAfeyn}.

  From the preceding discussions, and as is easily verified by explicit
calculation, the $Z^0$ mediated contribution to the moment distribution is
proportional to up to $\cos^2\!\theta$ only. It is not so obvious that this
is still the case for the $W^\pm$ mediated contribution and the
interference between the two contributions.

  The analytical expression for this process, which includes both the
$Z^0$ and $W^\pm$ contributions, is given in Eqns.~(2)--(5) of
Ref.~\cite{berends}. There is an error in equation (3) available therein,
and one should correct this error. The $s'$ in the overall factor should
only multiply the first two terms inside the square brackets. The
expression without this correction is dimensionally inconsistent.
  We have performed four more independent calculations, out of which two
are analytic and the other two use HELAS \cite{helas}, and verified that
the numbers are consistent with each other.

  To motivate the discussion, let us first consider the $W^\pm$ mediated
background in the limit of large $W^\pm$ mass. In this case the $W^\pm$
fusion graph (1) of Fig.~\ref{nunuAfeyn} vanishes and we have only
graphs (2) and (3), both of which can be written in terms of an effective 
contact interaction between two charged currents.

  However, by Fierz transformation, this contact interaction between two
charged currents is equivalent to the contact interaction between two
neutral currents, and so the amplitude has identical form to the $Z^0$
mediated background \cite{heavyW}. Hence the distribution is only
proportional to up to $\cos^2\!\theta$.

  Furthermore, also by Fierz transformation, the SUSY signal
$\widetilde{\gamma}\widetilde{\gamma}\gamma$, or more generally
$\widetilde{\chi}^0_1\widetilde{\chi}^0_1\gamma$, has the same structure
in the limit of large selectron mass. We thus obtain the expressions
listed in Tabs.~\ref{tab_prefac}, \ref{tab_coeffs}.

  It is found \cite{Montagna:1998ce} that at LEP energies, although the
above consideration of the heavy $W^\pm$ limit is clearly inappropriate,
the shapes of angular distributions corresponding to the $Z^\pm$ and
$W^\pm$ mediated contributions are fairly similar. There is no guarantee
that this will still hold at future LC energies, and explicit evaluation
indeed shows that there is nonvanishing $\cos4\theta$ and higher
contributions.

  Even so, what becomes clear is the following. Let us consider the general
case of the production of a photon and two invisible objects. If the
particle exchanged in the $t$ channel is heavy, as is the case for the SUSY
$\widetilde{\chi}^0_1\widetilde{\chi}^0_1\gamma$ production mediated by
$t$-channel selectron exchange, the interaction is approximated by a
contact interaction and then the distribution can only show a $\cos2\theta$
behaviour. On the other hand, if the exchanged particle is light, as is the
case for the $W^\pm$ contribution to the SM background at LC energies, the
interaction is expected to be well understood and so the contribution can
be subtracted, either from the bare distribution or from the Fourier
coefficients. This provides partial justification for our procedure.

  We also note that the $W^\pm$ mediated background involves left-handed 
electrons and right-handed positrons, and therefore it can be reduced by 
means of beam polarization.

 \subsection{Experimental Limitations}
 \label{experimentallimitations}

  The experimental limitations which we take into account are the
calorimeter resolution, angular coverage and beamstrahlung. The angular
resolution is not expected to be a serious constraint. Beamstrahlung is
treated using CIRCE \cite{CIRCE} as discussed in the next Subsection.

  For the EM calorimeter resolution, the EM track energy resolution $\Delta E$ 
is expected to be \cite{TESLATDR,JLCTDR}:
 \begin{eqnarray}
 \frac{\Delta E}{E} &=& \frac{15 \%}{\sqrt{E(\mathrm{GeV})}} \oplus 1 \%
 \ ({\rm JLC}) \\
                 &=& \frac{20 \%}{\sqrt{E(\mathrm{GeV})}} \oplus 0.6 \%
 \ ({\rm TESLA}) .
 \end{eqnarray}
 This smears the photon energy distribution. As an example, for a 100
GeV photon, $\Delta E$ at JLC is 1.8 GeV, wheres at TESLA $\Delta
E=2.1$ GeV. For photon energies lower than about 250 GeV, the energy
resolution is always better at JLC. As a pessimistic estimate, although
this choice does not affect our results in any visible way, let us adopt
the TESLA numbers in our analysis.

  Next, let us consider the coverage of the EM calorimeter. Typical
numbers for this coverage \cite{TESLATDR,JLCTDR} are in the range 50
mrad--200 mrad. In this study, let us adopt $|\cos\theta|<0.995$, which
corresponds to an angular cut-off of $\theta_{\rm min}=100$ mrad
\cite{TESLATDR}.

  The derivation of the Fourier coefficients is completely successful only
when the range of integration is propotional to the cyclic bound.
  The angular coverage is expected to become a problem if the Fourier 
component under consideration changes phase near this unmeasured region.
For a Fourier component proportional to $\cos(2n\theta)$, the first change 
of phase occurs at $\theta=\pi/4n$. If we wish to measure $n\leq3$, we 
need a coverage down to $\theta_{\rm min}<\pi/12=260$ mrad.

\begin{figure}[ht]
 \centerline{\epsfig{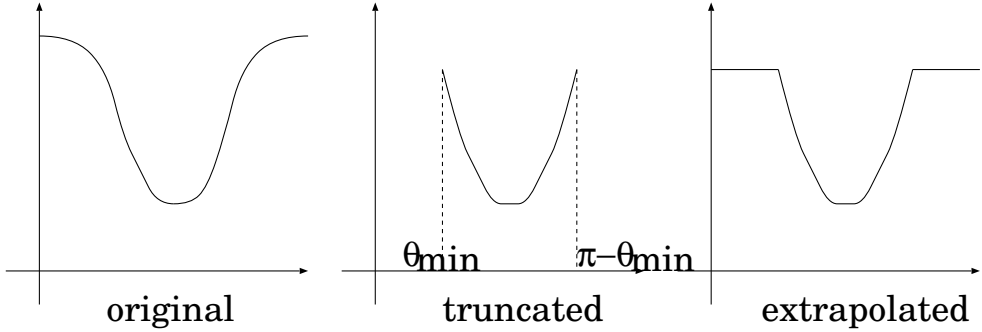}}
 \caption{Illustrating the extrapolation procedure. The vertical axis is
the $E_T^2$ moment distribution, and the horizontal axis is $\theta$. The
$x_\gamma$ dependence is not shown.}
 \label{extrapolation}
 \end{figure}

  Explicit evaluation shows that the vanishing of the term
$a_3\cos(6\theta)$ is indeed contaminated with the above coverage,
$|\cos\theta|<\cos\theta_{\rm min}=0.995$. In order to evade this problem,
as a simple remedy, we perform the extrapolation, that is, the
approximation that the values in the non-coverage regions have constant
values equal to the end-points. This is illustrated in
Fig.~\ref{extrapolation}.

The coefficient functions in Eqn.~(\ref{a0}) and (\ref{an})
are then redefined by
\begin{eqnarray}
a_0(x_\gamma) &=& \frac{1}{2\pi} \Biggl\{
\int_{2\theta_{\rm min}}^{2[\pi-\theta_{\rm min}]}
\frac{E_T^2 d^2 \sigma}{d\cos\theta dx_\gamma} d(2\theta)
\\ \nonumber
&+& \int_0^{2\theta_{\rm min}} \left.
\frac{E_T^2 d^2 \sigma}{d\cos\theta dx_\gamma}
\right|_{\theta=\theta_{\rm min}}
d(2\theta) 
+ \int_{2[\pi-\theta_{\rm min}]}^{2\pi} \left.
\frac{E_T^2 d^2 \sigma}{d\cos\theta dx_\gamma}
\right|_{\theta=\pi-\theta_{\rm min}} 
d(2\theta) \Biggr\},
\end{eqnarray}
\begin{eqnarray}
a_m(x_\gamma) &=& \frac{1}{\pi} \Biggl\{
\int_{2\theta_{\rm min}}^{2[\pi-\theta_{\rm min}]}
\frac{E_T^2 d^2 \sigma}{d\cos\theta dx_\gamma} \cos(2m\theta) d(2\theta)
\\ \nonumber
&+& \int_0^{2\theta_{\rm min}} \left.
\frac{E_T^2 d^2 \sigma}{d\cos\theta dx_\gamma}
\right|_{\theta=\theta_{\rm min}} 
\cos(2m\theta) d(2\theta) 
\\ \nonumber
&+& \int_{2[\pi-\theta_{\rm min}]}^{2\pi} \left.
\frac{E_T^2 d^2 \sigma}{d\cos\theta dx_\gamma}
\right|_{\theta=\pi-\theta_{\rm min}}
\cos(2m\theta) d(2\theta) \Biggr\} 
\qquad(m=1,2,3,\ldots).
\end{eqnarray}

  We have found that this remedy is sufficient to overcome the problem at
hand.
  We note that in the presence of ISR, the angles quoted above are
measured in the laboratory frame rather than the centre-of-mass frame.  
This makes the imposition of the above cut-and-extrapolation procedure
more difficult, and we adopt an approximate, `wrong', solution which is to
impose the angular cut and extrapolation in the centre-of-mass frame. Of
course the evaluation of the integral is otherwise performed in the
laboratory frame, as required.

  For the centre-of-mass energy we mainly take $\sqrt{s}=500$ GeV. We also
consider the $\sqrt{s}$ dependence by varying this to 1 TeV.  We omit the
discussion of statistics in this paper, but the luminosity may be
considered to be in the region 100--500 fb$^{-1}$ per year
\cite{TESLATDR,JLCTDR}.

 \subsection{Initial State Radiation}\label{sect_isr}

  The non-singular part of the higher order perturbative contributions is
expected to mainly affect the normalization and not the shape of the
distributions. However, the logarithmically enhanced bremsstrahlung
corrections must be considered for realistic simulation. There is another
source of ISR, that is, beamstrahlung. We consider both effects.

  Beamstrahlung is the radiation of photon caused by the interaction
between the incoming electron and positron bunches. This smears both the
centre-of-mass energy and the momentum along the beam axis. We use CIRCE
\cite{CIRCE} in order to estimate this effect. 

  Bremsstrahlung is the radiation caused by the interaction between the
electron and the positron participating in the annihilation event. We
estimate this effect by using the integrated expression of \cite{brems}.
This does not take into account the effect of large angle emission into
the detector, nor does it take into account the emission of more than one
photon from each incoming particle, but is expected to model the dominant
effect well.

  As an illustration, the typical energy carried by a bremsstrahlung photon is
\begin{eqnarray}
\int_{0}^{1} d(E,x) \ (E x) \ dx,
\label{bremsstrahlung}
\end{eqnarray}
where, from the expression of \cite{brems},
\begin{eqnarray}
d(E,x) \sim \frac{\alpha_{EM}}{\pi} \frac{2}{x} (\log \frac{E}{m_{e}}-
\frac{1}{2}) \sim \frac{0.063}{x} \ (E \sim 250\!\mathrm{GeV}).
\end{eqnarray}
  Hence Eqn.~(\ref{bremsstrahlung}) yields 16 GeV for a 250 GeV beam.  
This energy taken away by the photon affects both the centre-of-mass
energy and the frame. This change of frame implies a Lorentz boost to the
final state photon and so the angular distribution is also affected, as
well as the energy distribution.

  The subdominant `hard bremsstrahlung', where photons enter into the
detectable region, is not treated in this paper, but we expect that a
minimum $E_T$ cut is necessary to reduce such effect. Such a cut would also
be effective in reducing the background coming from low $E_T$ Bhabha and
forward photon events.
  This would be equivalent to assigning an energy dependent cut for the
minimum polar angle $\theta_{\rm min}$, and we expect that this can be
dealt with by the extrapolation procedure adopted for calorimeter coverage
in the preceding Subsection.
  As an example, for a 50 GeV photon, requiring $E_T>10$ GeV corresponds to
$\theta_{\rm min}=201$ mrad.
  Further study is desirable to quantify this effect.

\begin{figure}[ht]
 \begin{center}
 \epsfig{file=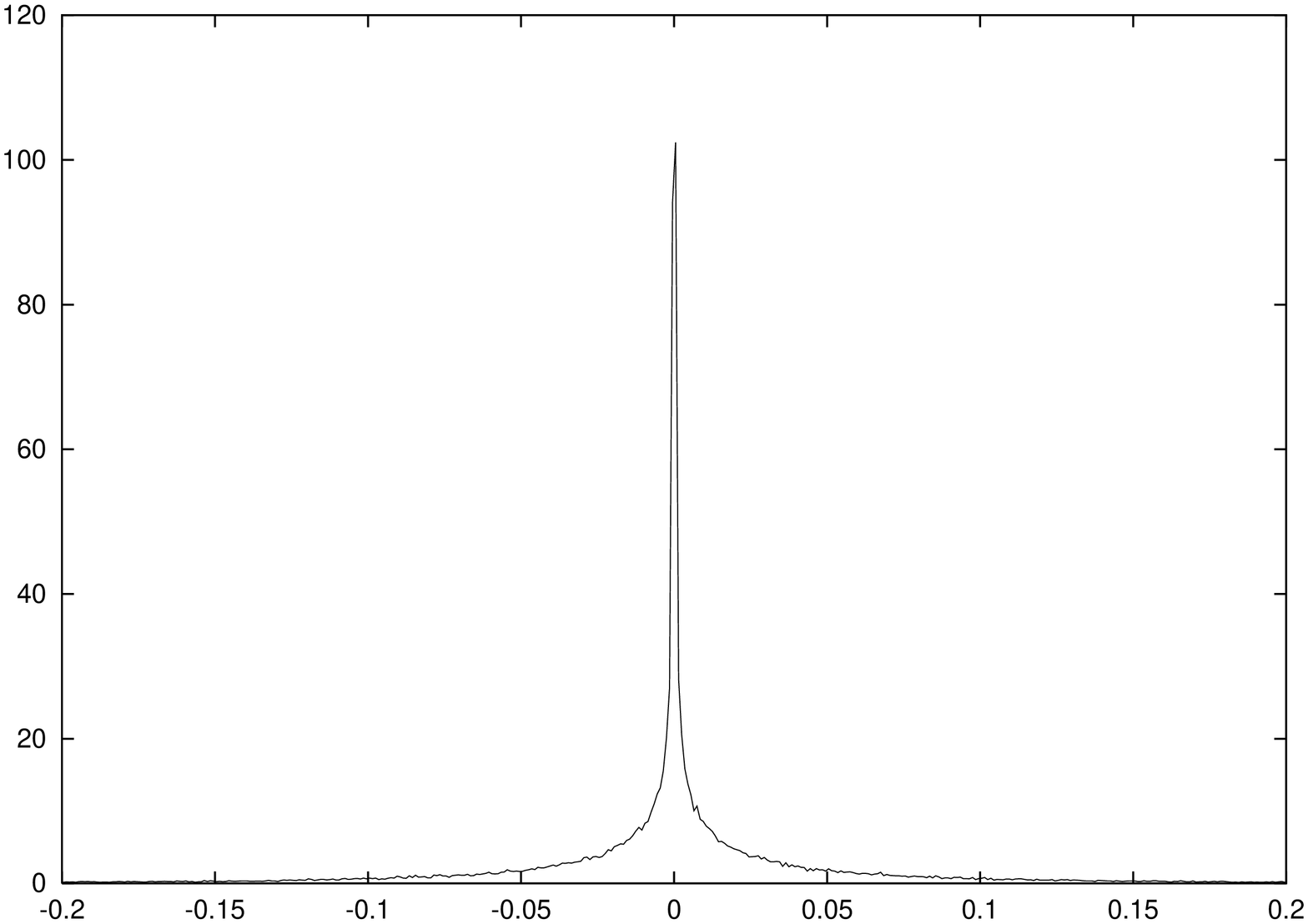,angle=270,width=7cm}
 \epsfig{file=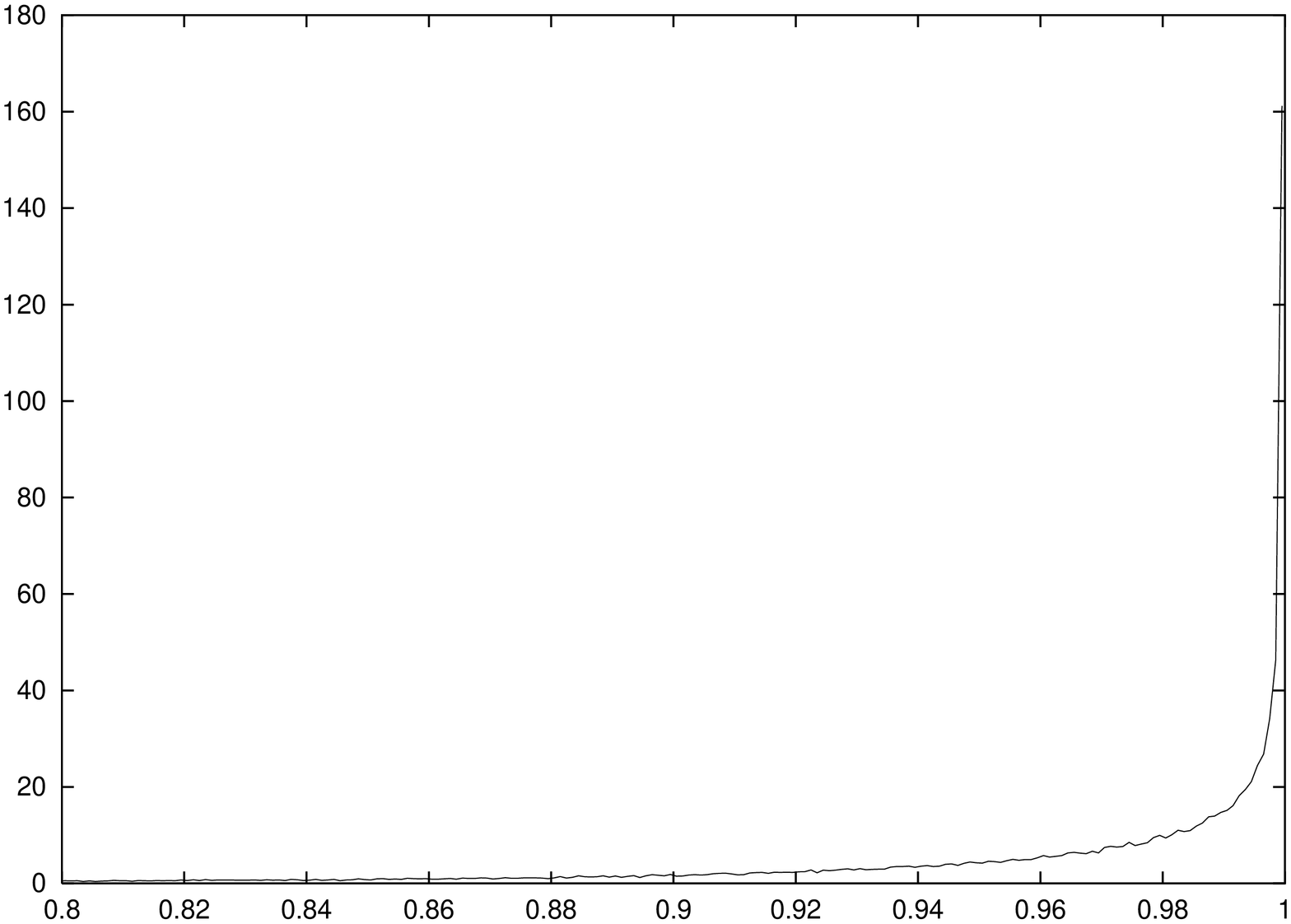,angle=270,width=7cm}
 \put(-410,5){\tiny $\frac2\sigma\frac{d\sigma}{d\cos\theta}$}
 \put(-200,5){\tiny $\frac2\sigma\frac{d\sigma}{d\cos\theta}$}
 \end{center}
 \caption{Distributions of $\cos\theta$ (left) and $x_\gamma$ (right) of
the $x_{\gamma\rm CM}=1, \cos\theta_{\rm CM}=0$ events after beamstrahlung
and bremsstrahlung. The distributions are normalized to unity.}
 \label{bremstest}
 \end{figure}

  As an illustrative example, we consider a configuration in which a
photon is produced at $x_{\gamma\rm CM}=1, \cos\theta_{\rm CM}=0$. The
Lorentz boost is easily shown to be given in terms of the electron and
positron momentum fractions $x_1$ and $x_2$ by:
 \begin{eqnarray}
 \cos\theta_{\rm lab} &=& \frac{x_1-x_2}{x_1+x_2}, \\
 x_{\gamma\rm lab}    &=& \frac{x_1+x_2}2.
 \end{eqnarray}
  The effective centre-of-mass energy is given by $\sqrt{\hat
s}=\sqrt{sx_1x_2}$.
  One can immediately verify that the transverse energy is unaffected by
this boost and is given by $E_T=\sqrt{\hat s}/2$. We can now plot these
functions for a given centre-of-mass energy $\sqrt{s}$ and beamstrahlung
spectrum. Taking $\sqrt{s}=500$ GeV and the TESLA beamstrahlung spectrum,
we obtain the results shown in Fig.~\ref{bremstest}.
  From these figures, we can obtain a rough estimate of the effect of ISR.  
$\cos\theta$, or equivalently $\theta$, is smeared typically by about 0.01
(radians). Although this smearing is small enough for us to expect that 
ISR will not change the angular distribution so much as to produce 
non-vanishing higher Fourier components, there is a relatively long tail 
to the distribution which may yield non-negligible effect.

  For small smearing, $x_1,x_2\sim1$, the polar angle dependence of
the angular smearing can be shown to go as:
 \begin{equation}
 \delta\theta_{\rm lab}\approx\frac{x_1-x_2}2\sin\theta.
 \end{equation}
  As expected, the angular smearing is the largest when $\theta=\pi/2$.

  The energy fraction is smeared by about 0.01, which is comparable to the
calorimeter resolution. This is again accompanied by a long tail. As the
cross section is proportional to $(\sqrt{s})^{n}$, we expect that this
suppression in the effective centre-of-mass energy may lead to a
significant suppression of the cross section.

 \section{Results}\label{results}

  In this Section, we follow our analysis numerically. The parameters
related to the graviton emission, the fundamental scale and the number of
extra dimensions, are taken to be $M_D = 1$ TeV and $n = 2$ for now.

\begin{figure}[ht]
 \begin{center}
 \epsfig{file=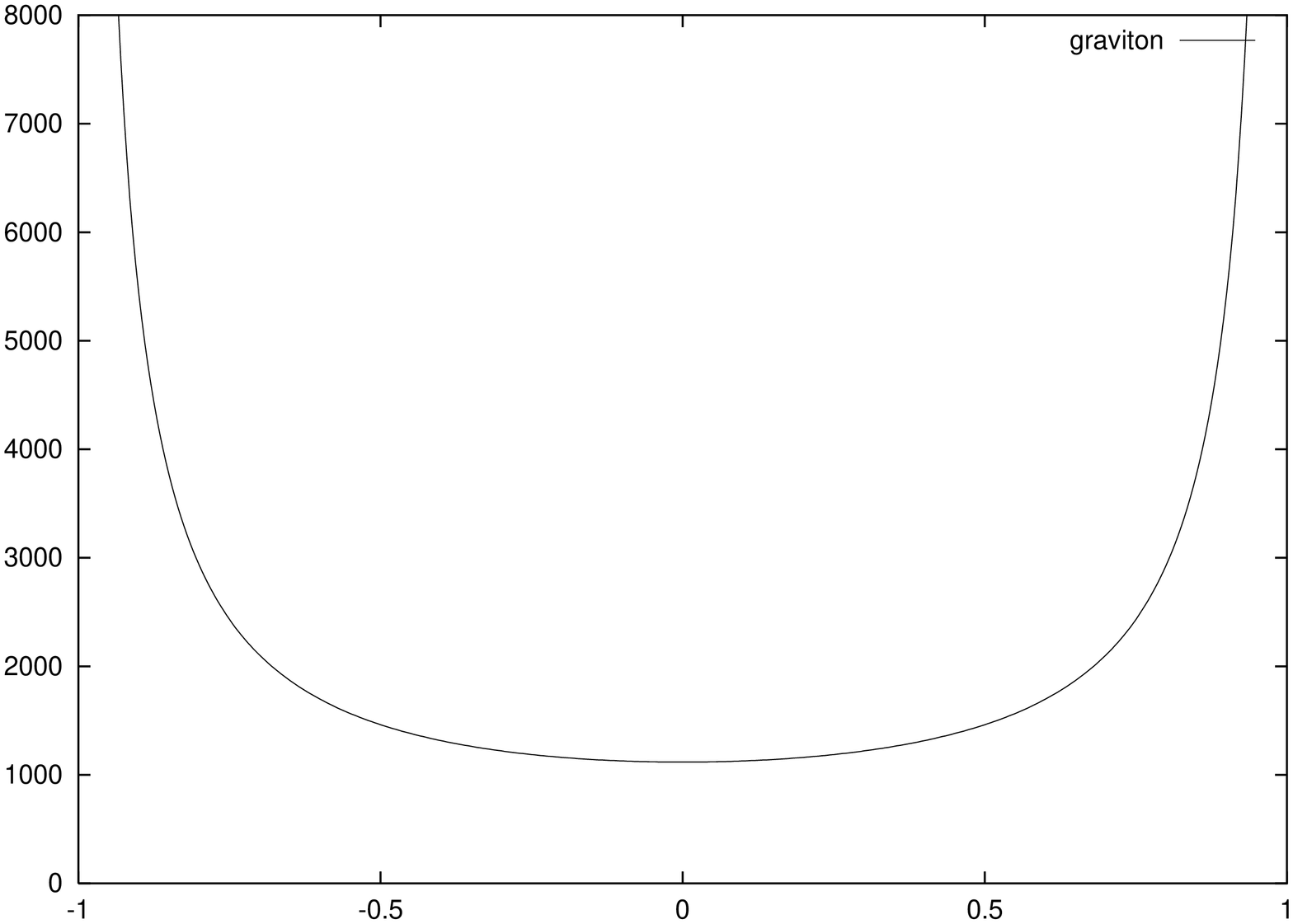,angle=270,width=7cm}
 \epsfig{file=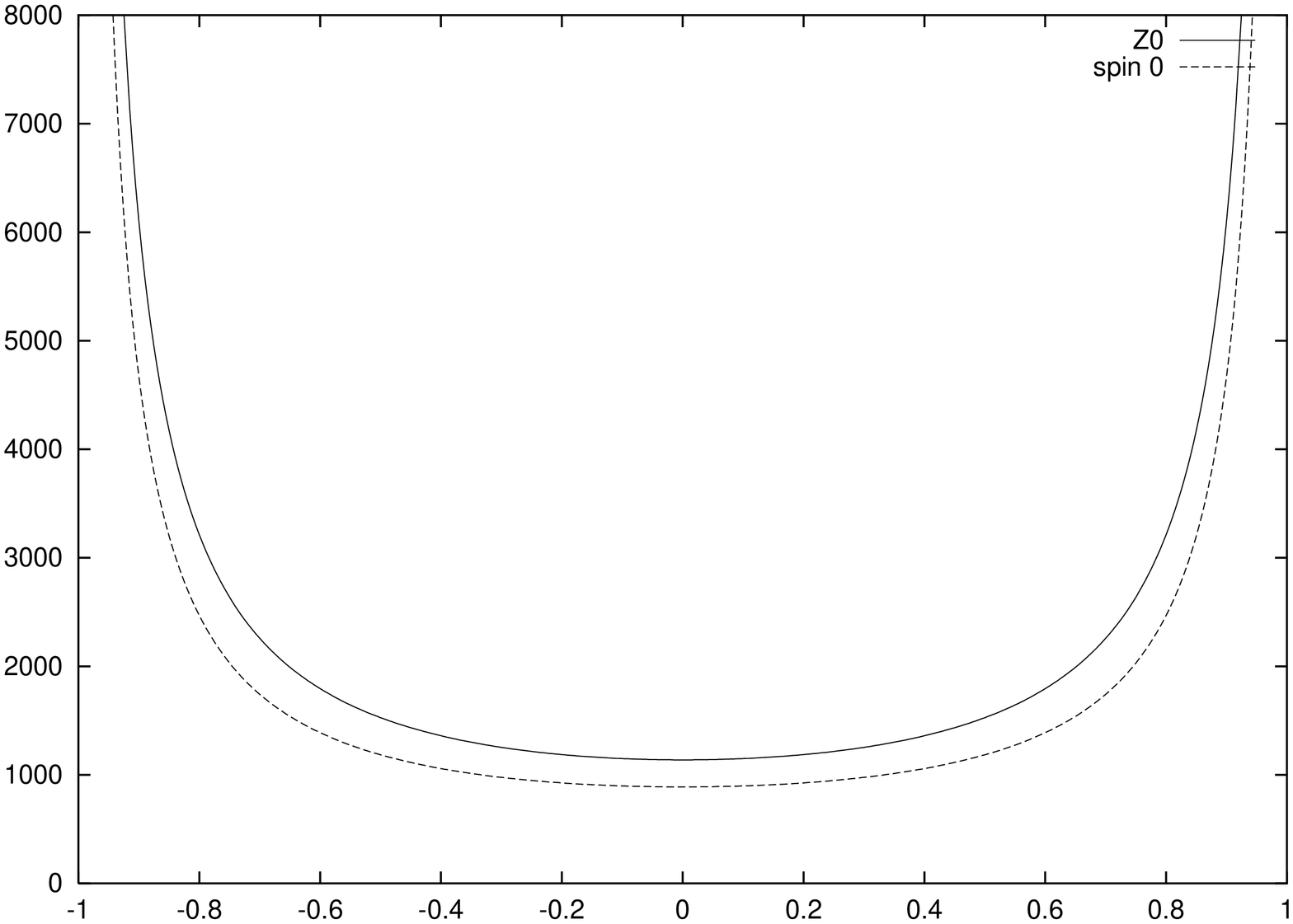,angle=270,width=7cm}
 \put(-305,-142){\tiny $\cos\theta$}
 \put(-100,-142){\tiny $\cos\theta$}
 \put(-410,5){\tiny $\frac{d\sigma}{d\cos\theta}$\ /fb}
 \put(-200,5){\tiny $\frac{d\sigma}{d\cos\theta}$\ /fb}
 \end{center}
 \caption{The subprocess level $\cos\theta$ distribution, for the graviton
events integrated over $0.05<x_\gamma<1$ (left), and the on-shell
$Z^0\to\nu\bar\nu$ events (right). For the $Z^0$ events, we also show the
hypothetical production of a spin-0 object of the same mass (lower curve).
The normalization is arbitrary in this case.}
 \label{ang-dis}
 \end{figure}

  In Fig.~\ref{ang-dis}, the $\cos\theta$ dependence for the process $e^+
e^- \rightarrow G \gamma$ at $\sqrt{s}=500$~GeV is shown, together with the
corresponding plot for on-shell $Z^0$ production. A cut on the energy
fraction of the photon is taken as $x_\gamma > 0.05$. In the plot for
on-shell $Z^0$ production, we also show the distribution for the production
of a hypothetical scalar $Z^0$. The normalization is arbitrary in this
case.
  For this set of parameters, we see that the total rates are quite
similar.

  From Fig.~\ref{ang-dis}, we see that at first sight, there is no clear
distinction between the distributions due to the production of objects of
different spin. We therefore look mainly at the $E_T^2$ moment
distribution, as explained in the preceding Sections.

  We then apply the Fourier expansion to this moment distribution, as
expressed in Eqns.~(\ref{a0}) and (\ref{an}). As mentioned before, this is
not the only method to extract the Fourier coefficients, and it may turn
out that a $\chi^2$ based fit is more practical. We take this approach
because it is the most theoretically elegant, and also because it provides
a convenient way of parameterizing the two dimensional distribution of
$x_\gamma$ and $\cos\theta$.

\begin{figure}[hbt]
 \begin{center}
 \epsfig{file=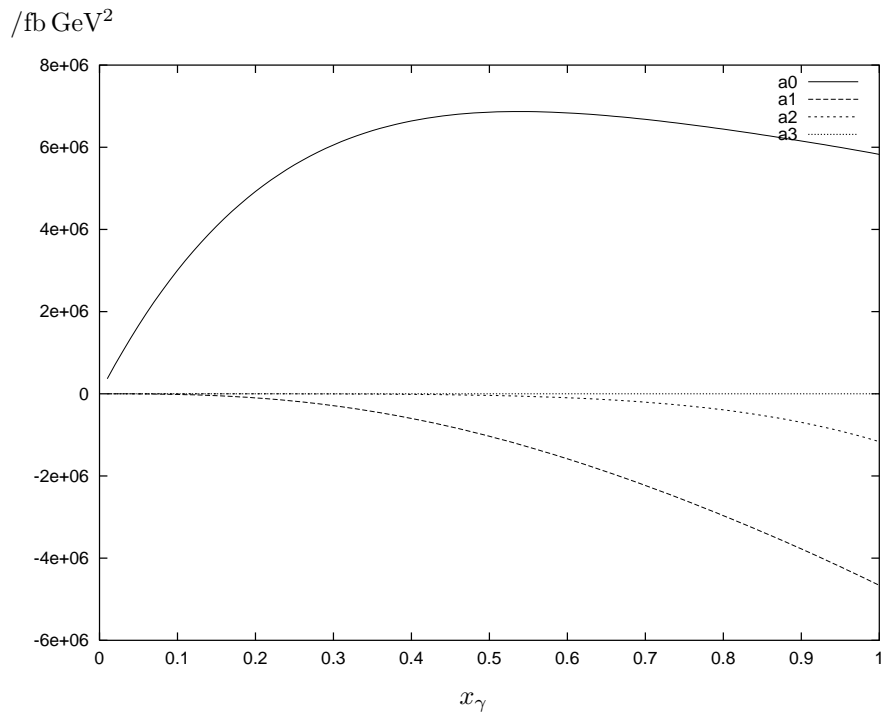,angle=270,width=12cm}
 \put(-340,5){\small /fb\,GeV$^2$}
 \put(-170,-250){\small $x_\gamma$}
 \end{center}
 \caption{The Fourier coefficients in units of fb\,GeV$^2$ for graviton
events, before applying any corrections.}
 \label{bare-coef}
 \end{figure}

  Fig.~\ref{bare-coef} shows the Fourier coefficients for graviton events,
before applying any corrections, as functions of the energy fraction
$x_\gamma$. We see that only the first three coefficient functions are
non-zero, as expected.

\begin{figure}[hbt]
 \begin{center}
 \epsfig{file=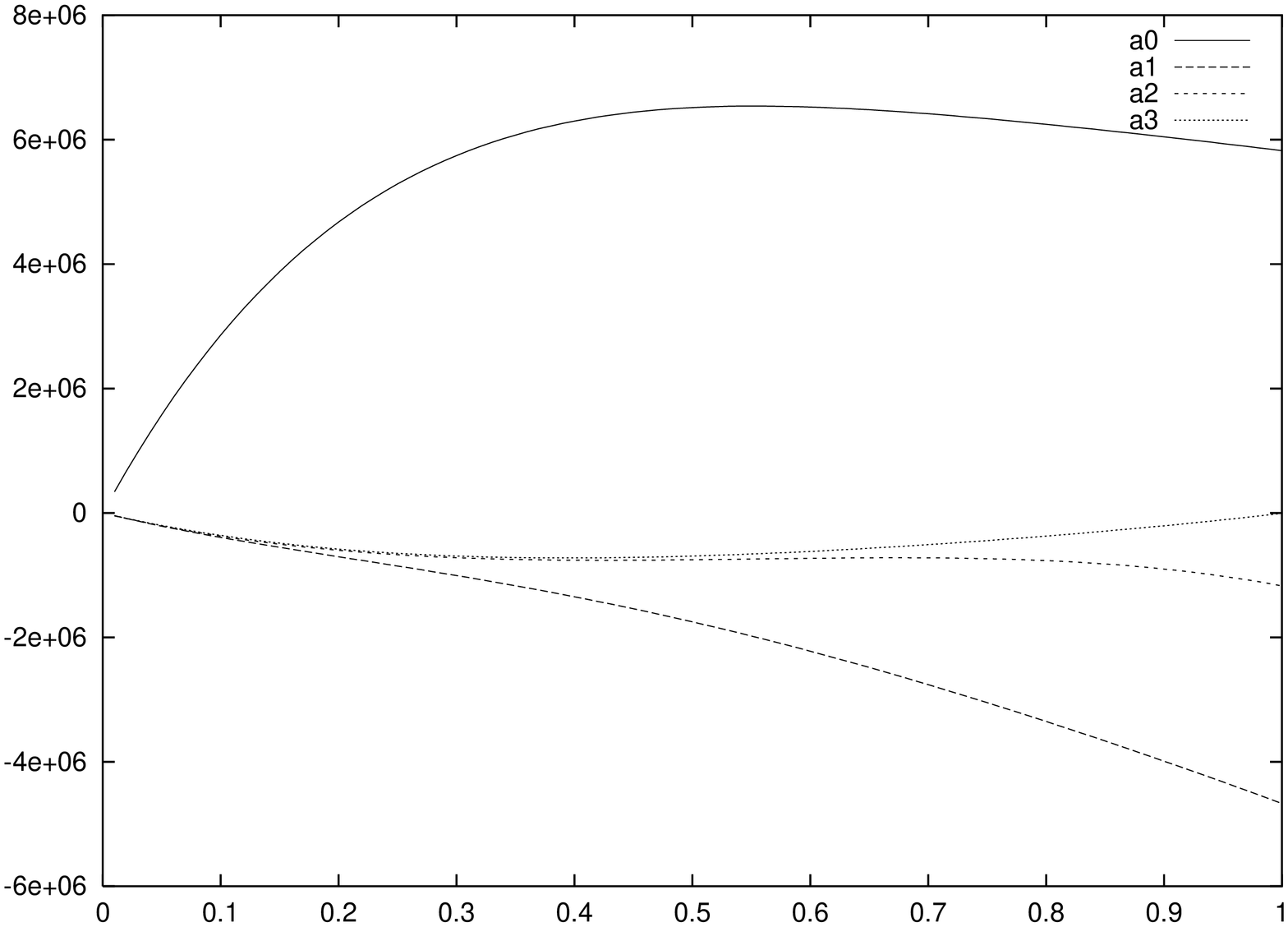,angle=270,width=7cm}
 \epsfig{file=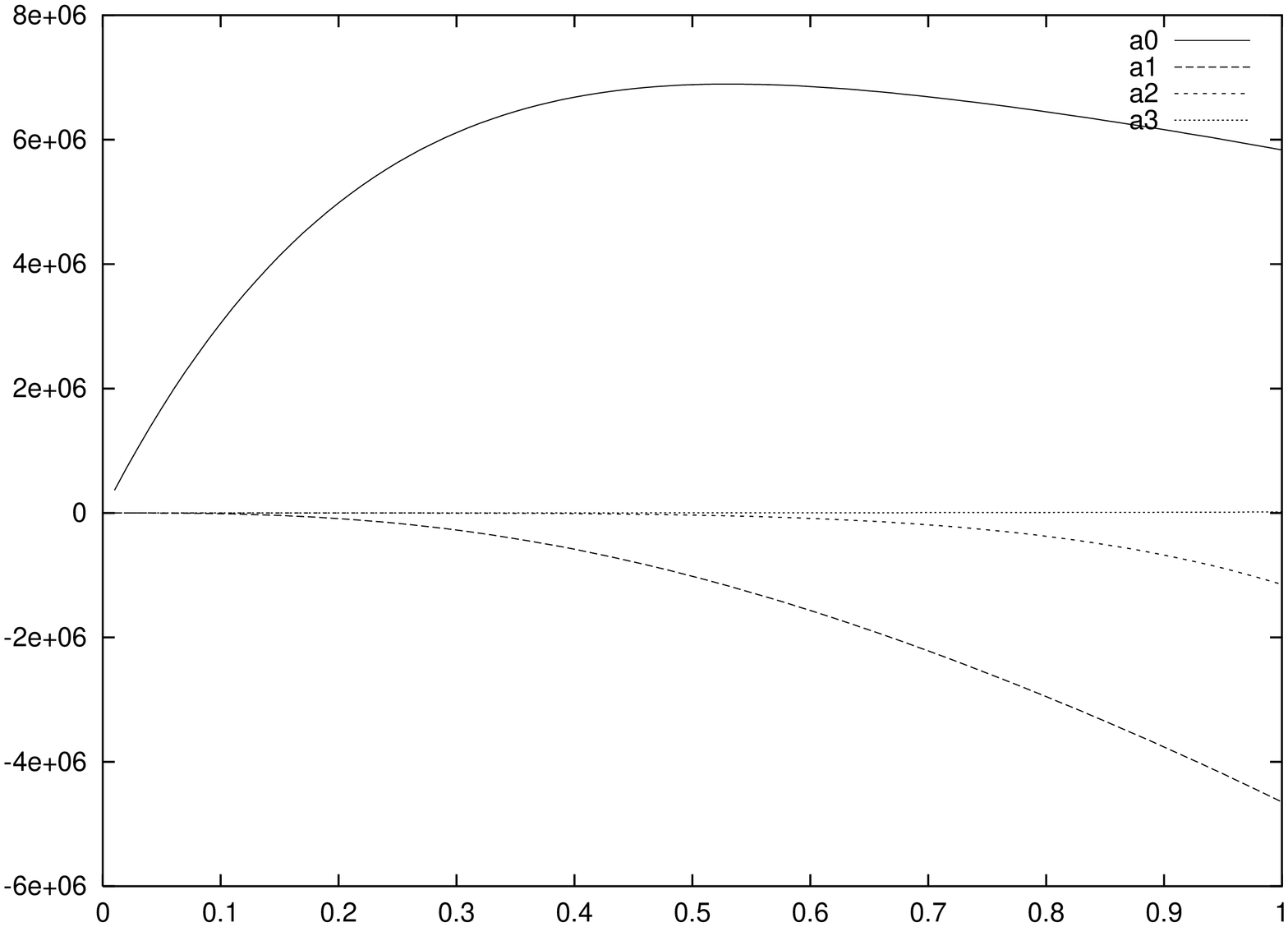,angle=270,width=7cm}
 \put(-305,-142){\tiny $x_\gamma$}
 \put(-95,-142){\tiny $x_\gamma$}
 \put(-410,5){\tiny /fb\,GeV$^2$}
 \put(-200,5){\tiny /fb\,GeV$^2$}
 \end{center}
 \caption{The Fourier coefficients for graviton events, after including
the calorimeter coverage (left), and after the correction by extrapolation 
(right).}
 \label{ang-coef}
 \end{figure}

  When we truncate the distribution to account for the finite coverage of
the calorimeter as stated in Sect.~\ref{experimentallimitations}, $a_m$ for
large $m$ acquires an artefactual contribution and becomes non-zero.
However, we recover the original distribution by the extrapolation
procedure proposed above. The situation is illustrated in
Fig.~\ref{ang-coef}.

  As the corrected distribution shown in Fig.~\ref{ang-coef} looks almost
identical to the original distribution shown in Fig.~\ref{bare-coef}, we 
conclude that the procedure is sufficient for our purpose.

  Let us now take into account the effects of bremsstrahlung and
beamstrahlung and, perhaps less importantly, the calorimeter energy
resolution. For the beamstrahlung and calorimeter energy resolution,
although the choice is arbitrary and leads to little difference in the
result, let us adopt the TESLA parameters.

\begin{figure}[ht]
 \begin{center}
 \epsfig{file=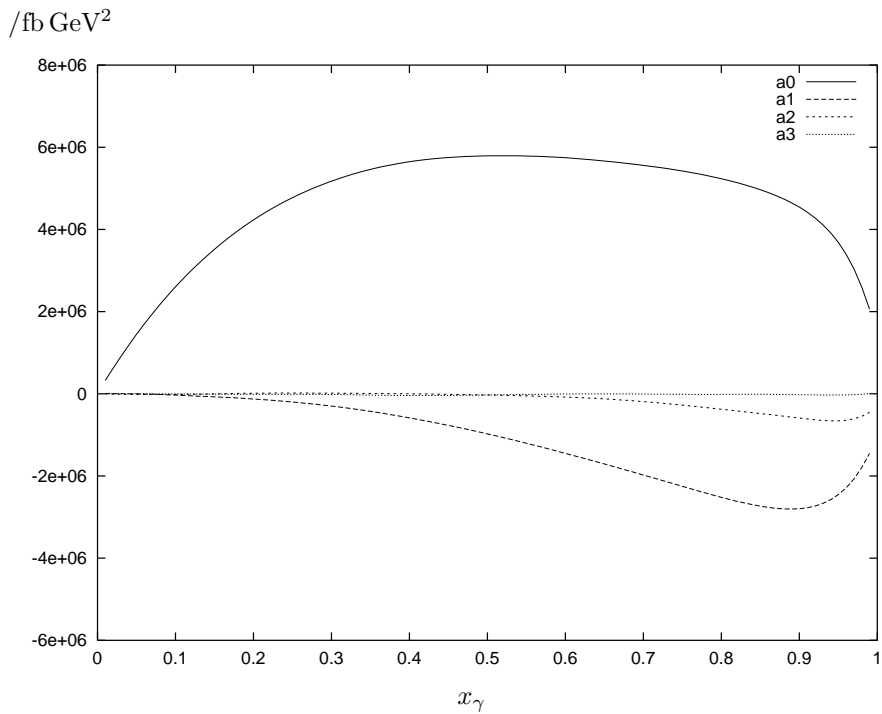,angle=270,width=12cm}
 \put(-340,5){\small /fb\,GeV$^2$}
 \put(-170,-250){\small $x_\gamma$}
 \end{center}
 \caption{The Fourier coefficients for graviton events, including all
corrections.}
 \label{final-coef}
 \end{figure}

  The effect of the ISR is explained and illustrated in
Sect.~\ref{sect_isr}. Applying the same ISR spectrum to the signal, we
obtain the results shown in Fig.~\ref{final-coef}. We see that the
coefficients have dropped significantly, and there is further suppression
as $x_\gamma\to1$. This is as expected.
 The coefficient $a_3$ remains zero even with the inclusion of ISR.

\begin{figure}[ht]
 \begin{center}
 \epsfig{file=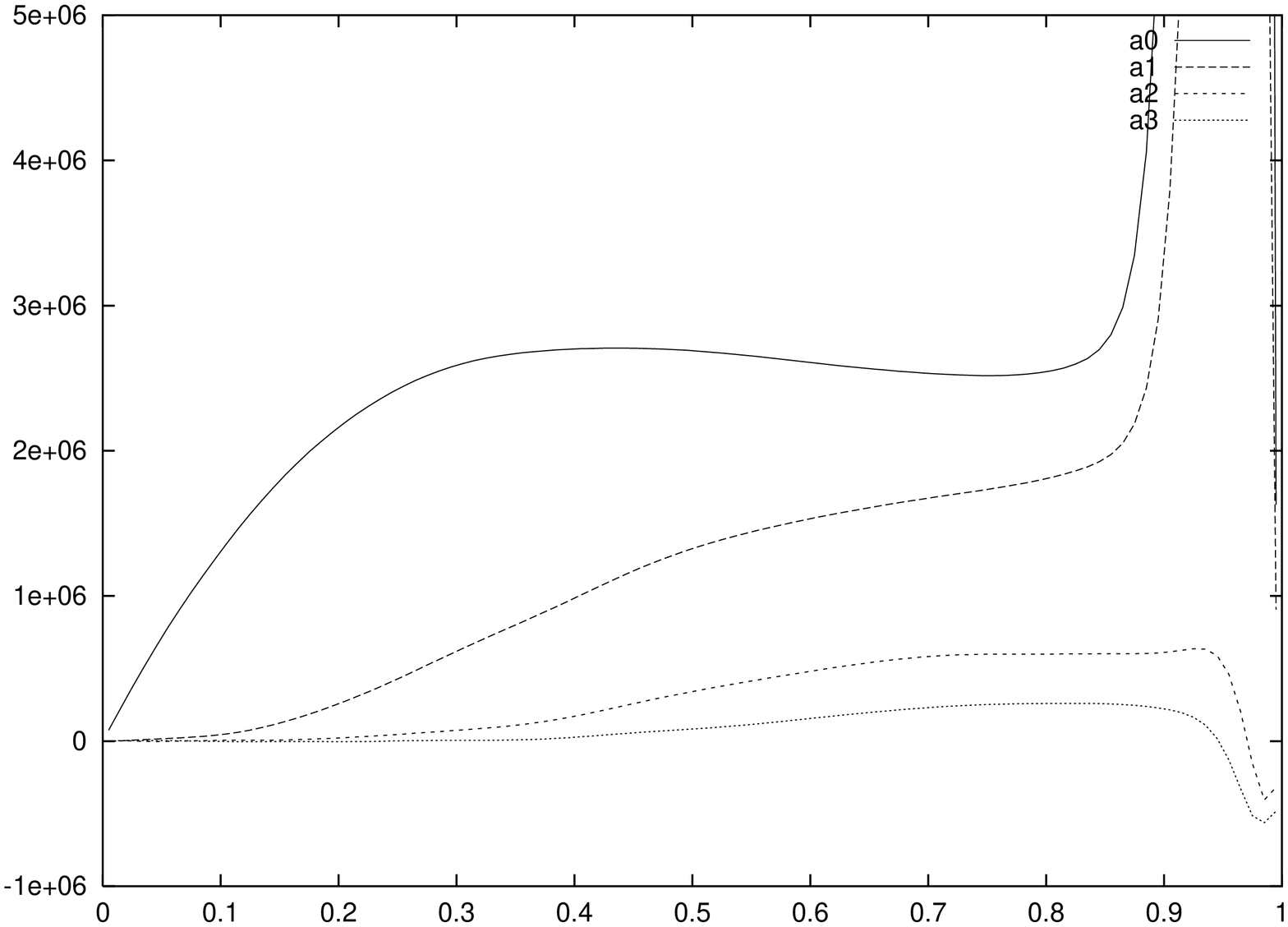,angle=270,width=7cm}
 \epsfig{file=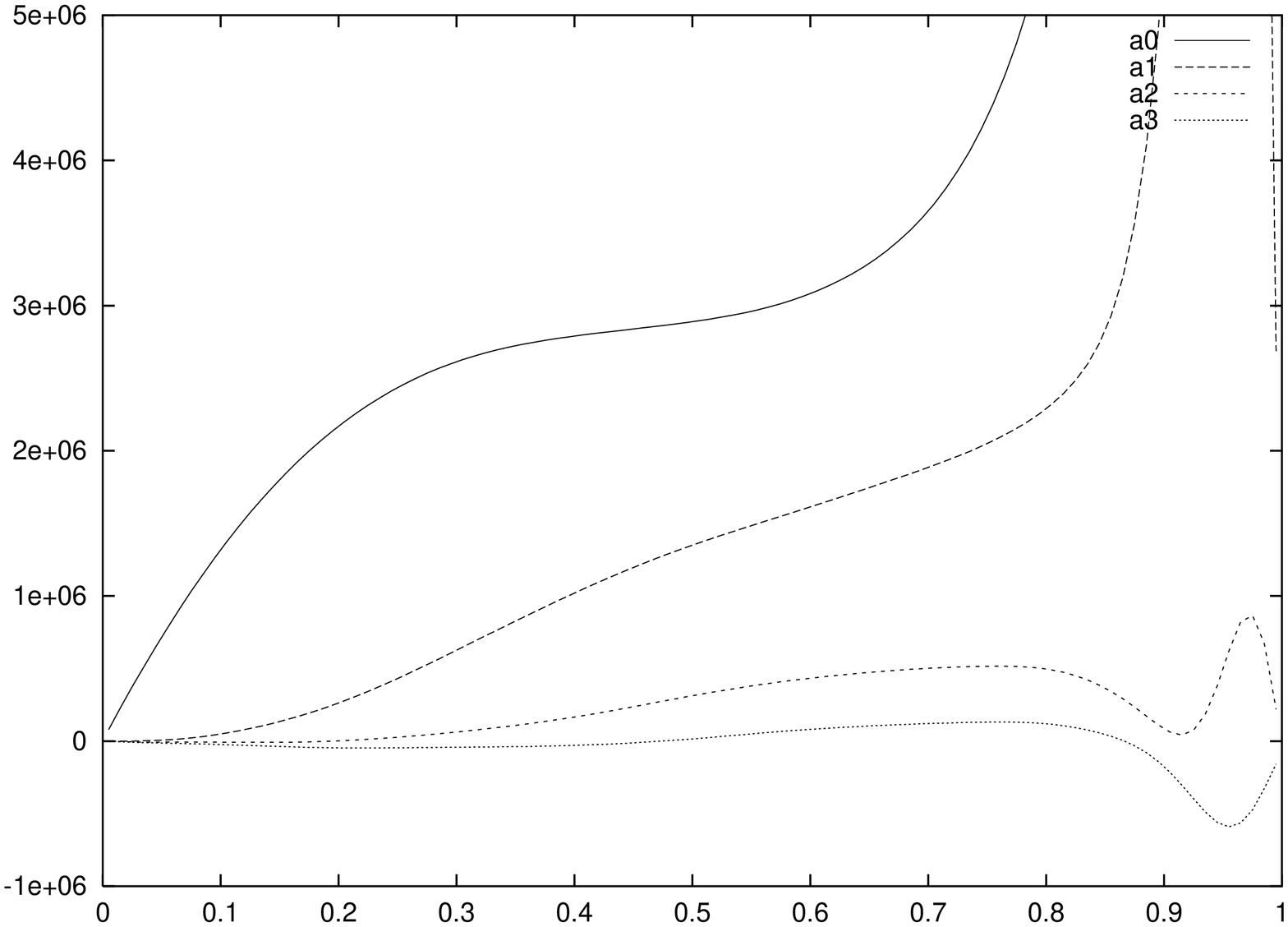,angle=270,width=7cm}
 \put(-305,-142){\tiny $x_\gamma$}
 \put(-95,-142){\tiny $x_\gamma$}
 \put(-410,5){\tiny /fb\,GeV$^2$}
 \put(-200,5){\tiny /fb\,GeV$^2$}
 \end{center}
 \caption{The Fourier coefficients for the SM continuum background, 
without the experimental and bremsstrahlung corrections (left), and 
including all corrections (right).}
 \label{bkgd_coef}
 \end{figure}

  Let us now turn our attention to the SM continuum background. The
distribution is shown in Fig.~\ref{bkgd_coef}.
  Again, the rate is quite similar to the signal for this set of extra
dimensional parameters.
  The dominant part of the background distribution for $x_\gamma<0.8$ is
due to the $W^\pm$ mediated contribution. The rise of the cross section
near $x_\gamma=1$ is due to the $Z^0$. The coefficients $a_2$ and $a_3$ do
not have a peak for this reason. However, there is the interference term
between the $Z^0$ and $W^\pm$ mediated contributions, and this interference
term gives rise to the observed behaviour near $x_\gamma=0.95$.
  With the inclusion of ISR, the $Z^0$ peak broadens. At the same time,
the angular smearing due to the ISR gives rise to additional contributions
to $a_2$ and $a_3$ at the $Z^0$ peak, which lead to the interesting 
behaviour observed near that region.

  From these figures, it is clear that the proper simulation of the ISR is
essential to understanding both the signal and the background.

  As noted in Sect.~\ref{sect_bkgd}, $a_2$ and higher coefficients are in
general non-zero for $\sqrt{s}$ sufficiently higher than $M_W$. The
explicit evaluation, as shown in Fig.~\ref{bkgd_coef}, confirms this.  
Although these coefficients are non-zero, we note, again, that the shape
of the background is sufficiently well understood that the distribution
can be subtracted away, either from the bare distribution or the Fourier
coefficients.

  Finally, let us turn our attention to the centre-of-mass energy,
$\sqrt{s}$, dependence of the distributions, and the dependence of the
signal distribution on the number of extra dimensions $n$.

\begin{figure}[ht]
 \begin{center}
 \epsfig{file=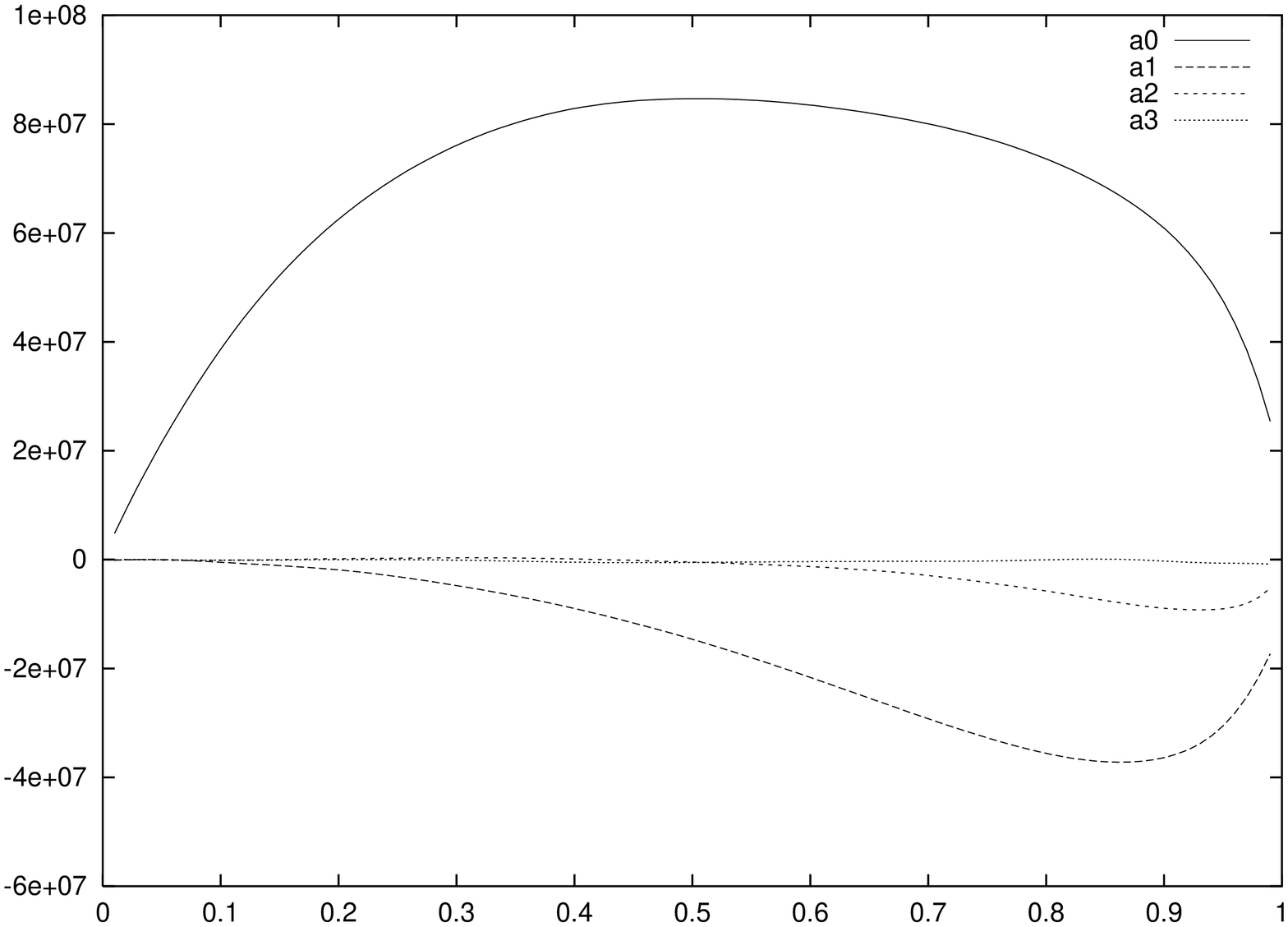,angle=270,width=7cm}
 \epsfig{file=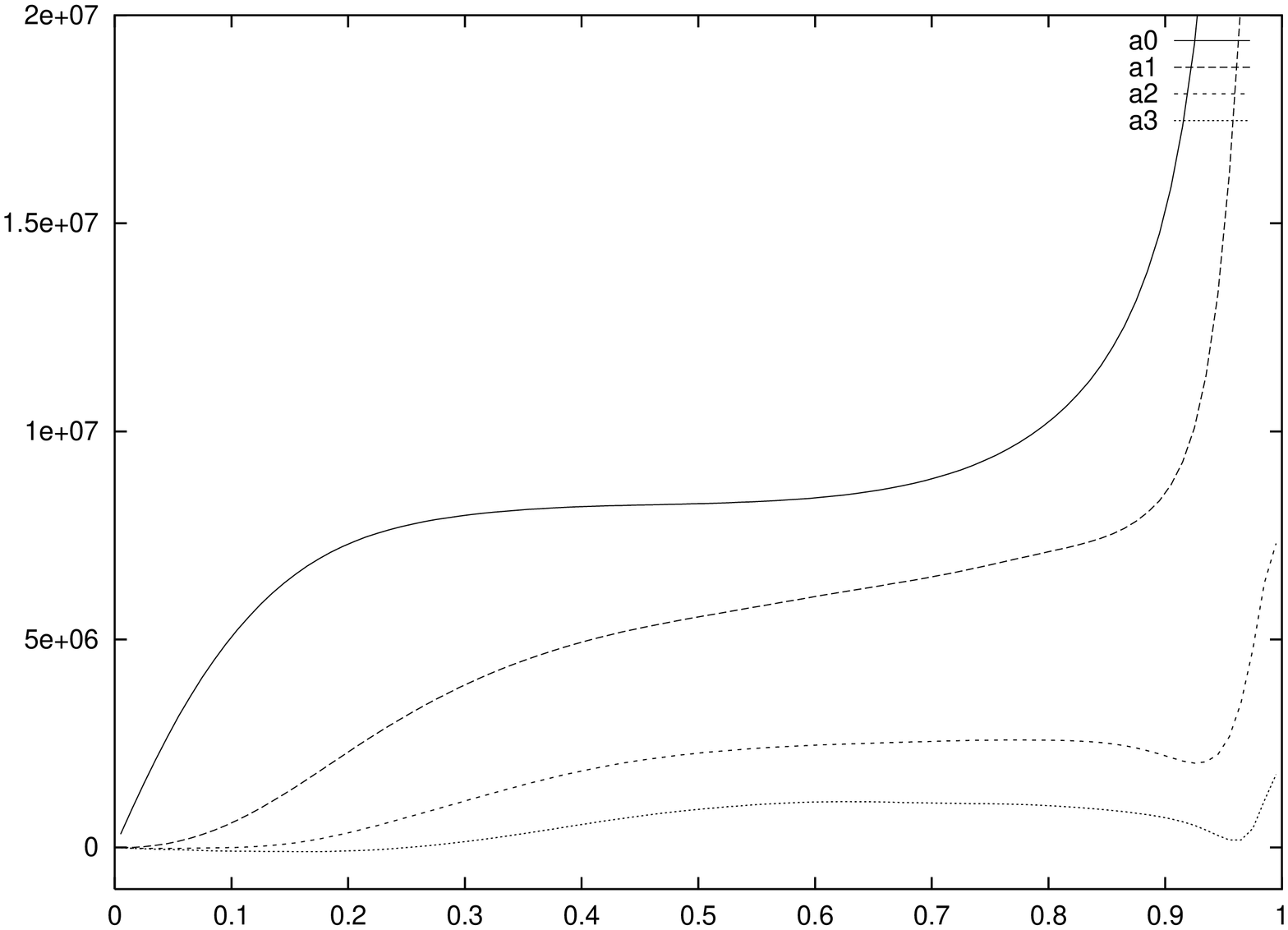,angle=270,width=7cm}
 \put(-305,-142){\tiny $x_\gamma$}
 \put(-95,-142){\tiny $x_\gamma$}
 \put(-410,5){\tiny /fb\,GeV$^2$}
 \put(-200,5){\tiny /fb\,GeV$^2$}
 \end{center}
 \caption{The signal (left) and background (right) Fourier coefficients, 
at $\sqrt{s}=1$ TeV, including all corrections.}
 \label{roots_coef}
 \end{figure}

  In order to understand the centre-of-mass energy dependence, we
calculated the signal and background Fourier coefficients, which have so
far been calculated for $\sqrt{s}=500$ GeV, for $\sqrt{s}=1$ TeV, including
all corrections. The result is shown in Fig.~\ref{roots_coef}.

  We see that there is a significant change in the shape of the background
distribution when the centre-of-mass energy is raised. The signal
distribution does not change much, but if there is any difference, it is
expected to be due to the increased beamstrahlung. The total rate is
increased for the signal and reduced for the background. For the case of
the background, although the cross section is reduced, the size of the
$E_T^2$ moment distribution is increased because of the centre-of-mass
energy dependence of the total effective $e^-e^+$ charged current cross 
section which, in the limit of large energy, becomes flat.

\begin{figure}[ht]
 \begin{center}
 \epsfig{file=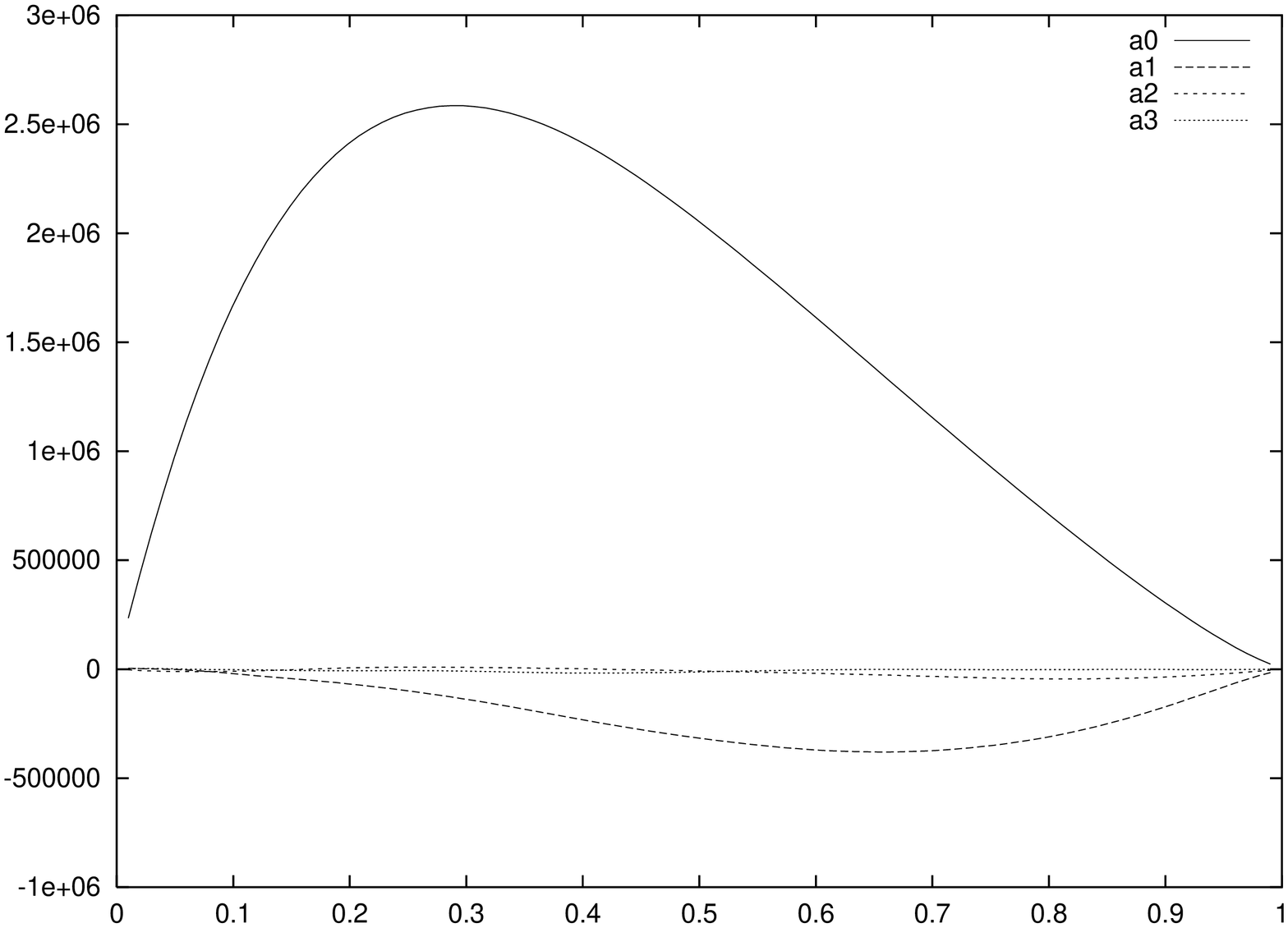,angle=270,width=7cm}
 \epsfig{file=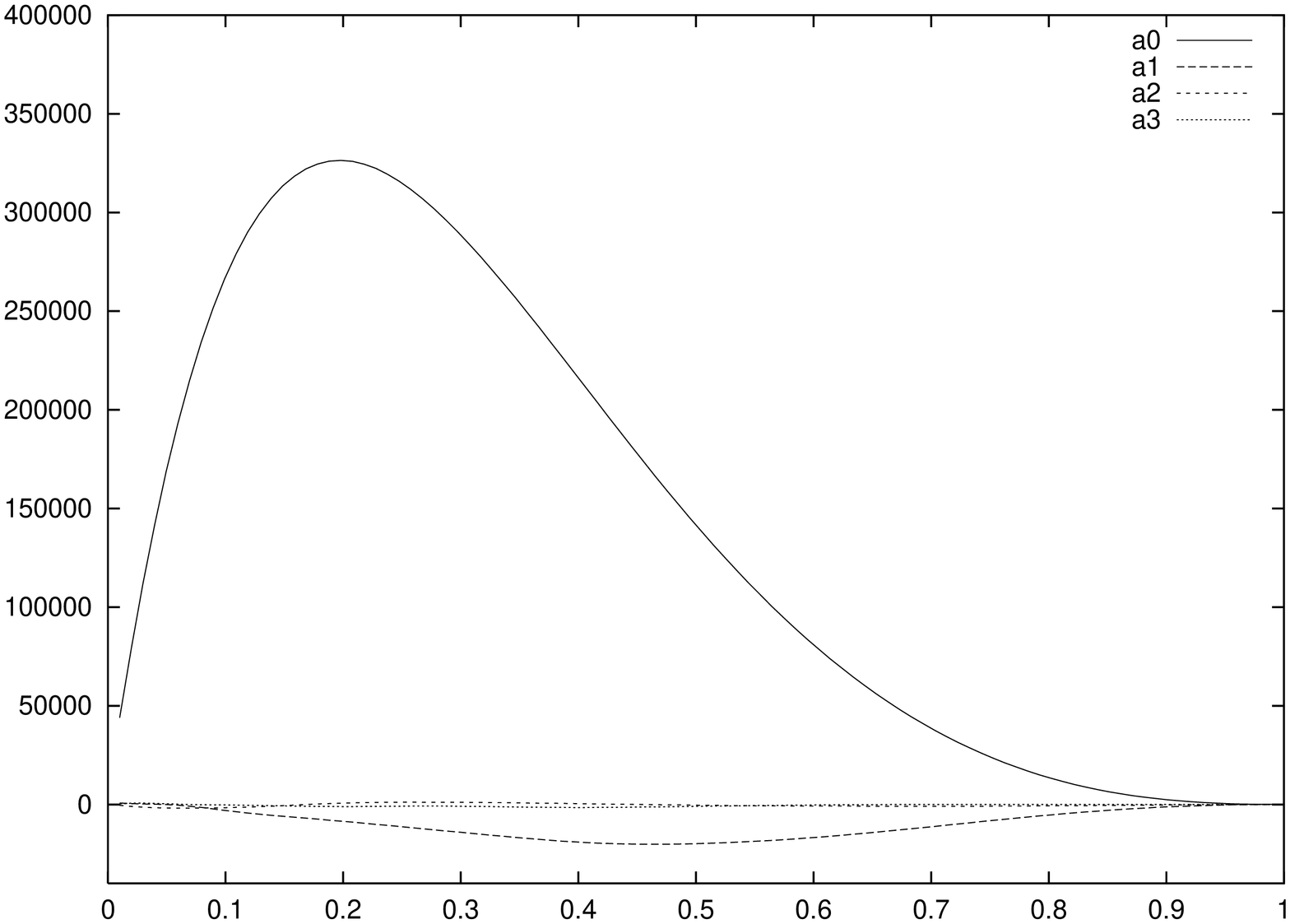,angle=270,width=7cm}
 \put(-305,-142){\tiny $x_\gamma$}
 \put(-95,-142){\tiny $x_\gamma$}
 \put(-410,5){\tiny /fb\,GeV$^2$}
 \put(-200,5){\tiny /fb\,GeV$^2$}
 \end{center}
 \caption{The signal Fourier coefficients for $n=4$ (left) and $n=7$ 
(right).}
 \label{noed_coef}
 \end{figure}

  In order to understand the dependence on the number of extra dimensions
$n$, we calculated the signal Fourier coefficients for $n=4$ and $n=7$, at
$\sqrt{s}=500$ GeV, again including all corrections. The result is shown in
Fig.~\ref{noed_coef}.

  We see that as $n$ becomes larger, the Fourier coefficients are more
suppressed towards larger $x_\gamma$ because of the factor
$(1-x_\gamma)^{n/2-1}$. The higher Fourier coefficients, which rise with
$x_\gamma$ as powers of $x_\gamma$, become suppressed, and the $E_T^2$
moment distribution becomes more flat. In this limit, it becomes impossible
to determine the spin from the angular distribution only, and one would
have to resort to the $\chi^2$ fitting, for example, of the leading Fourier
coefficient $a_0$ as a function of $x_\gamma$ in order to discriminate the
extra dimensional hypothesis from other hypotheses. This method would also
be useful when there is not sufficient statistics to utilize the angular
distribution.


  It may be argued that the presence of a quartic polynomial in the
coefficient $a_0$, as shown in Tab.~\ref{tab_coeffs} indicates the spin-2
nature. This is less rigourous, and there would remain the possibility
that the shape is due to the distribution of gravitons. We also note that
the size of the quartic term is not large.

  To some extent the difference in the shape of the Fourier coefficients
can be exploited to `measure' the number of extra dimensions, or more
precisely the distribution of the Kaluza-Klein modes. One can alternatively
measure the $\sqrt{s}$ dependence of these coefficients, proportional to
$\sqrt{s}^{n+2}$ for $n$ extra dimensions, to yield the same information.

  Having outlined the application of our procedure thus, one may like to
question the statistical efficacy of our observables. Although a formal
discussion would require a more detailed study, it is a simple matter to
have a rough estimation of the statistics. If we assume a relatively low
integrated luminosity $\mathcal{L}= 100$ fb$^{-1}$, and look, for
instance, at the numbers at 500 GeV shown in Fig.~\ref{final-coef}, and
take $0.8<x_\gamma<0.9$, say, a quick estimation of the numbers show that
the significance defined as `$S/\sqrt{B}$'$=\sqrt{\mathcal{L}\Delta
x_\gamma a_2^2/(E_T^2a_0)}\sim2$ so that 95\% confidence level can be
attained even with low luminosity for this set of $n$ and $M_D$.

 \section{Conclusions}\label{conclusions}

  We have proposed a way of measuring the `spin' of invisible objects,
including gravitons, produced at future LC.

  Our method is to measure the Fourier coefficients $a_m(x_\gamma)$
($m=0,1,2\cdots$) of the $E_T^2$ moment distributions. The presence of
terms up to $a_S$ and the absence of the higher coefficients indicates the
spin-$S$ nature of the invisible object.

  We have provided a theoretical justification for this observation, and
have carried out a simulation based on this principle. Our simulation
includes the consideration of beamstrahlung, bremsstrahlung, calorimeter
coverage, calorimeter resolution and the background, and we believe that
the simulation is realistic. The ISR, composed of beamstrahlung and
bremsstrahlung, has a particularly marked effect on both the signal and
the background. Our study is nevertheless incomplete, as the statistical
study and the error estimation have not been performed.
  When doing so, it is desirable that further attention is paid to the
effect of the hard/multiple bremsstrahlung contribution which must be
dealt with more carefully.

  This method as stated is useful when the number of extra dimensions $n$
is 2. If we believe in the astrophysical constraint, this case may be
ruled out. For greater $n$, the direct application of this procedure is
not suitable because the higher Fourier coefficients become small. Even in
that case, the method provides a parameterization of the cross section
which separates the angular and energy dependence. The resulting
$x_\gamma$ dependent coefficient $a_0$, for example, can be subjected to
$\chi^2$ analysis in order to test the extra dimensional hypothesis.

  We also note that the shape of the Fourier coefficient functions will
provide a measure of the `number' of extra dimensions. More generally, the
measurement of spin aside, what our method provides is a framework by which
we may parameterize single photon events at future LC.

 \section*{Acknowledgements}

  We thank Francesca Borzumati, Keisuke Fujii, Kazuo Fujikawa, Sachio
Komamiya, Jae-Sik Lee, Yasuhiro Okada, Masahiro Yamaguchi and Tsutomu
Yanagida for helpful comments and discussions. In particular, we owe a
great deal to the insight provided by Keisuke Fujii for the more
experimental aspects.

\end{document}